\documentclass[10pt,twocolumn,english,prl,showpacs,superscriptaddress]{revtex4}
\usepackage[latin9]{inputenc}
\usepackage{textcomp}
\usepackage{amsmath}
\usepackage{amssymb}
\usepackage{graphicx}

\makeatletter
\@ifundefined{textcolor}{}
{%
 \definecolor{BLACK}{gray}{0}
 \definecolor{WHITE}{gray}{1}
 \definecolor{RED}{rgb}{1,0,0}
 \definecolor{GREEN}{rgb}{0,1,0}
 \definecolor{BLUE}{rgb}{0,0,1}
 \definecolor{CYAN}{cmyk}{1,0,0,0}
 \definecolor{MAGENTA}{cmyk}{0,1,0,0}
 \definecolor{YELLOW}{cmyk}{0,0,1,0}
}

\usepackage{epstopdf}
\usepackage{color}
\makeatother

\usepackage{babel}
\begin{document}

%%%% Article title to be placed here
\title{Introduction: Localized Structures in Dissipative Media: From Optics to Plant Ecology}

\author{%%%% Author details
M. Tlidi$^{1}$, K. Staliunas$^{2}$, K. Panajotov$^{3,4}$, A.G. Vladimirov$^{5}$ and M.G. Clerc$^{6}$}

%%%%%%%%% Insert author address here
\address{$^{1}$ D\'{e}partment de Physique, Universit\'{e} Libre de Bruxelles (U.L.B.), CP\ 231, Campus Plaine, B-1050 Bruxelles, Belgium.\\
$^{2}$Departament de F{\'i}sica i Enginyeria Nuclear, Universitat Polit{\`{e}}cnica de Catalunya, Colom 11, 08222 Terrassa (Barcelona), Spain.\\
$^{3}$ Brussels Photonics Team, Department of Applied Physics and Photonics
(B-PHOT TONA), Vrije Unversiteit Brussels, Pleinlaan 2, B-1050 Brussels, Belgium.\\
$^{4}$Institute of Solid State Physics,72 Tzarigradsko Chaussee Blvd., 1784 Sofia, Bulgaria\\
$^{5}$Weierstrass Institute, Mohrenstrasse 39, D-10117 Berlin, Germany.\\
$^{6}$Departamento de F{\'i}sica, FCFM, Universidad de Chile, Blanco Encalada 2008, Santiago, Chile.}

%%%% Abstract text to be placed here %%%%%%%%%%%%
\begin{abstract}
Localised structures in dissipative appears in various fields of
natural science such as biology, chemistry, plant ecology, optics
and laser physics. The proposed theme issue is to gather  specialists from various fields of non-linear
science toward a cross-fertilisation among active areas of research. This is a 
cross-disciplinary area of research dominated by the nonlinear optics
due to potential applications for all-optical control of light, optical storage, and information processing.  This theme issue contains contributions from 18 active groups involved in localized structures field and have all made significant contributions in recent years.
\end{abstract}
%%%%%%%%%%%%%%%%%%%%%%%%%%%

%%%%%%%%%% Insert the texts which can accomdate on firstpage in the tag "fmtext" %%%%%

\maketitle
\section{Introduction}
Localized structures are "dissipative structures" found far from equilibrium. The term "dissipative structure" has been introduced by Ilya Prigogine to describe the spontaneous appearance of periodic structures as a result of irreversible processes maintained by exchanges of matter and/or energy with a non-equilibrium environment \cite{Prigogine,Lefever}. Dissipative structures generally evolve on macroscopic scales and can only be maintained by continuous application of a non-equilibrium constraint. The spontaneous emergence of dissipative structures arise from a principle of self-organization that can be either in space and/or in time. A classic example of spatial self-organization is provided in the context of chemecal raction diffusion system referred to as Turing instability \cite{Turing}. This instability is characterized by a symmetry breaking that leads spontaneousely to the formation of periodic structures with an intrinsic wavelength. In this case the wavelength is determind solely by the dynamical parameters such as diffusion coefficient and the inverse characteristic time associated with chemical kinetics.  The emergence of dissipative structures with an extrinsic wavelength has been reported in an early report by  B{\'e}nard \cite{Benard}. Transition from conductive to convective regimes occurs in a plane horizontal layer of fluid heated from below, in which the fluid develops a regular pattern of convection cells refrered to as B{\'e}nard cells. In this case, the wavelength, i.e., the size of the pattern is determined by the thickness of the fluid layer and not by the dynamical parameters. The first experimental evidence of dissipative structures with an interinsic wavelength has been given by Castets et al.\cite{DeKepper} and by Ouyang and Swinney \cite{Swinney} by using CIMA (Chlorite-Iodide-Malonic-Acid) reaction.  Earlier, the Soviet scientist Belouzov has predicted  the oscillating behavior and the ability of chemical waves concentration to spread \cite{BZ1,BZ2}. However, dissipative structures have not been discussed. 

Spontaneous symmetry breaking and self-organisation phenomena appear in various field of nonlinear science such as nonlinear optics, fiber optics, plant ecology , fluid mecanics, granular matter, plant ecology. The link between the classical Turing instability and trasverse patterns in nonlinear optics has been first established in the seminal paper by Lugiato and Lefever \cite{LL}. Dissipative structures could be localized in space and/or in time. Many driven systems exhibit localized structures (LSs), often called localized spots and localized patterns, which may be either isolated, randomly distributed, or self-organized in clusters forming a well-defined spatial pattern.  Localized structures (LSs) are homoclinic solutions (solitary or stationary pulses) of partial differential equations. The conditions under which LSs and periodic patterns appear are closely related. Typically, when the Turing instability becomes sub-critical, there exists a pinning domain where localized structures are stable.  This is a universal  phenomenon  and a well documented issue in various fields of nonlinear science, such as chemistry, plant ecology,  and optics.  The experimental observation of LSs in driven nonlinear optical cavities  has motivated further the interest in this field of research. In particular, LSs could be used as bits for information storage and processing. Several overvious have published on this active area of research \cite{Rev1,Rev2,Mandel1997,Rev3,Rev4,Rev5,Rev6,Rev7,Rev8,Rev8-1,Rev9,Rev10,Rev11,Rev12,Rev13,Rev14,Rev15,Rev16,Rev17,Rev18,Rev19}

Collectively, the papers in this Special Issue provide an overview of recent advances in the formation of Localized structures in dissipative media. This special issue includes a selection of papers presented at the conference of "Dynamics days", Madrid 2013.  It concerns various topics of localized structures in dissipative media which are briefly  discussed below.

\section{Localized structures and pattern formation in nonlinear optics}
Before the invention of lasers, the optics was linear (with some small exceptions, like Raman scattering), and no nonlinear dynamical regimes were ever considered in optics. As the spontaneous pattern formation can occur in strongly nonlinear systems driven far away from equilibrium, no pattern formation was expected in optics. First ideas on the spatial pattern formation in optics appeared in early 70-ties, decade after the laser invention, when the relation between coherent laser field in laser resonators and fluids/superfluids was recognized, based on the derivation of complex Ginzburg-Landau equation for the light fields \cite{k1}. Also filamentation of strong light fields in self-focusing media (in analogy to singularity appearance in plasmas) \cite{k2} can be considered as the very initial works on the nonlinear light pattern formation. In spite of these initial ideas the real pattern forming in optics was considered unrealistic, mainly because of the weakness of nonlinearities there.  The real interest in nonlinear optical patterns started in the late 80-ies and early 90-ies. In \cite{k2-1,LL,k3}, a nontrivial nonlinear transverse mode formation in laser beams were shown. In \cite{k4}, it was recognized that the nonlinear laser equations (Maxwell-Bloch equations) admit vortex solutions, which were soon confirmed experimentally \cite{k5}. These pioneering works were followed by an increasing number of investigations, mostly attempting to understand the mechanisms of nonlinear pattern formation by deriving reduced models for a laser and other broad aperture nonlinear resonators, such as modified Ginzburg-Landau equation \cite{k6}, and Swift-Hohenberg equation \cite{k7,k8,k8-1,k8-2,k8-3}. Such approach has been applied for other types of nonlinear resonators filled with photorefractive crystal \cite{k9}, with parametric nonlinearity \cite{k10,k11}, also passive driven nonlinear-Kerr media \cite{Mandel1997,k13}. The number of works on different aspects of transverse pattern formation was exploding in 90-ies (see  \cite{Rev1,Rev2,Mandel1997,Rev3,Rev4}). Very different issues have been considered, which are impossible to review in this short introduction. To mention a few, three-dimensional patterns in resonators (localized light bullets) \cite{k47,k48,k48-1,k48-2,k48-3,k48-4,k49,k50}, patterns in "rocked" resonators \cite{k51,k52,k52-1}, and hyperbolic patterns \cite{k53,k54}. Exotic phenomena, like optical vortices and vortex ensembles \cite{k4,k5,k20,k21,k22,k23,k24}, which can show chaotic dynamics (vortex mediated turbulence) \cite{k24,k25}, and vortices that arrange themselves in regular lattices \cite{k26,k27,k28,k28-1,k28-2} have been demontrated. Nonlinear patterns also appear in other optical configurations such as: mirrorless configuration with unidirectional light propagation in nonlinear media, leading to filamentation effects, to formation of bright solitons \cite{k14}, and optical vortices \cite{k15,k16}; counterpropagating configuration leading to hexagons \cite{k17}; systems with feedback loop, showing a large variety of extended patterns \cite{k18,k19}. Localized structures were reported in bistable nonlinear resonators such as lasers with saturable absorbers) \cite{k29,Rev2,k31}, in passive nonlinear resonators \cite{k32,Scroggie}, and optical parametric oscillators \cite{k33,k34}. Phase solitons have been demonstrated far from any pattern forming instability \cite{k40,k40-1,k41,k42}.

\section{Localized structures in VCSELs}
Owing to their large Fresnel number and short cavity, Vertical-Cavity Surface-Emitting Lasers (VCSELs) are best suited for cavity soliton (CS) studies and potential applications (for recent reviews see \cite{Barbay_aot11,Rev14}). VCSELs were first suggested and realized in 1979 \cite{Soda_jjap79} and it took more than 10 years to bring them to comparable performance to edge-emitting lasers \cite{Iga03,VCSELs13}. Nowadays VCSELs are replacing edge-emitting lasers in short and medium distance optical communication links thanks to their inherent advantages: much smaller dimensions, circular beam shape that facilitates coupling to optical fibers, two-dimensional array integration and on wafer testing that brings down the production cost \cite{VCSELs13}. As VCSELs emit light perpendicular to the surface and the active quantum wells, their cavity length is of the order of 1 $\mu$m - the wavelength of the generated light. Thanks to the maturity of the semiconductor technology VCSELs can be made homogeneous over a size of hundreds of $\mu$ms while the characteristic CS size is about 10 $\mu$m. Furthermore, the timescales of the semiconductor laser dynamics and CS formation are in the ns scale, which allows for fast and accurate gathering of data. Finally, VCSEL physics and dynamics are quite well understood \cite{VCSELs13,Panajotov_jstqe13,Panajotov_jqe09}. Therefore, many theoretical and experimental studies on CS formation in VCSELs have been carried out \cite{Taranenko_pra00}-\cite{Elsass_apb10}. CSs have been successfully demonstrated in broad area VCSELs both below \cite{Taranenko_pra00,Barland_n02} and above \cite{Hachair_jstqe06} the lasing threshold when injecting a holding beam of appropriate frequency. Spatially localized structure has also been found out in medium size VCSELs but only by using their particular polarization properties \cite{Hachair_pra09}. CS lasers (CSLs) in a VCSEL system without a holding beam have been demonstrated both experimentally \cite{Tanguy_prl08} and theoretically \cite{Paulau_pre08} in VCSELs subject to frequency selective optical feedback and in face to face coupled VCSELs \cite{Genevet_prl08,Columbo_ejpd10}. In these systems, the VCSELs are placed in self imaging optical systems with either an external grating or another VCSEL biased below lasing threshold, so that the system becomes bistable. Lasing spots spontaneously appear in these systems and can be switched on and off by another laser beam. As a matter of fact broad area laser with saturable absorber has been the first system in which CS have been predicted and studied theoretically \cite{Rosanov_os88,Vladimirov_job99,Fedorov_pre00}. CSs in a monolithic optically pumped VCSEL with a saturable absorber have been demonstrated in \cite{Elsass} and their switching dynamics studied in \cite{Elsass_apb10}. Several applications of CSs in VCSELs have been demonstrated: optical memory \cite{Pedaci_apl06}, optical delay line \cite{Pedaci_apl08} and optical microscopy \cite{Pedaci_apl08b}.
\section{Localized structures in photonic crystals, left-handed materials and excitons-polaritons}
High-speed switching i.e., processing speed in optical systems combined with possibility of parallel processing in broad area systems gives a clear advantage of the optical systems with respect to electronic circuits. However, the optical systems have also a serious disadvantage with respect to electronic systems - the relatively large spatial extent of several wavelengths, i.e. of tens of micrometers. This is a serious disadvantage limiting application of nonlinear optical patterns in microprocessing technologies. Therefore, the straighforward application  of cavity solitons for optical memory and all-optical processing has never come to practice. Nevertheless, the advance of micro- and nano- fabrication technologies offers new possibilities for optical pattern: novel micro-modulated materials, like photonic crystals or nanomaterials that allow the manipulation of light propagation (zero or negative refraction and diffraction), thus suppressing the diffractive broadening of the beams, and strongly reducing the size of the nonlinear photonic structures \cite{k58,k59,k59-1}. On the other hand, the combined light-mater structures, the so called excitons and surface polaritons, are also of drastically reduced dimensions, due to light-matter coupling \cite{k60,k61,k62}. 

The fundamental limitations of LSs is their spatial size, which is due to the diffraction phenomenon in optical resonators. To overcome this limit, use of left-handed metamaterials is proposed, i.e. engineered materials with simultaneously negative permittivity and permeability \cite{Veselago,Shelby,Wiltshire,Shalev,Boardman}. These materials were first used by the scientific community to realize imaging systems with sub-wavelength resolution then to target potential applications including invisibility cloaks and perfect optical concentrators. They also allow for phase compensation in optical structures containing both left- and right-handed materials. Recently, it has been shown that such materials could be used to improve telecommunication systems by lowering power threshold and increasing data density in optical memories \cite{Kockaert06,Kockaert09,Tassin07,Gelens07}. More recently, an exprimental evidence of localized structure in left-handed materials has been provided \cite{Kozyrev}.

\section{Front dynamics in spatially extended systems}
Macroscopic systems under the influence of injection and dissipation of quantities such as energy and momenta usually exhibit coexistence of different states, which is termed multistability. This is clearly a demonstration that macroscopic systems are of nonlinear nature. Heterogeneous initial conditions usually caused by the inherent fluctuations generate spatial domains, which are separated by their respective interfaces. These interfaces are known as front solutions or interfaces or domain walls, etc. \cite{Rev4}. Interfaces between these metastable states appear in the form of propagating fronts and give rise to a rich spatiotemporal dynamics \cite{Pomeau,Langer,Collet}. Front dynamics occurs in systems as different as walls separated magnetic domains \cite{Eschenfelder}, directed solidification process \cite{Borzsonyi}, nonlinear optical systems \cite{Clerc2001,Gomilla2001,Clerc2004,Residori2005}, oscillating chemical reactions \cite{Petrov97}, fluidized granular media \cite{Aronson,Melo,Douady,Moon2001,Moon2003,Macias,Macias}, or population dynamics \cite{Fisher,Rev8,Clerc2005}.  From the point of view of dynamical system theory, in one spatial dimension a front is a nonlinear solution that is identified in the comoving frame system as a heteroclinic orbit linking two spatially extended states \cite{vanSaarloos,Coullet2002}.
The dynamics of the interface depends on the nature of the states that are connected. In the case of a front connecting a stable and an unstable state, it is called a Fisher-Kolmogorov-Petrosvky-Piskunov (FKPP) front  \cite{Rev8,Kolmogorov,VanSaarloos03}. One of the characteristic features of these fronts is that the speed they move is not unique, nonetheless determined by the initial conditions. When the initial condition is bounded, after a transient, two counter propagative fronts with the minimum asymptotic speed emerge \cite{Rev8,VanSaarloos03}. In case that the nonlinearities are weak, this minimum speed is determined by the linear or marginal-stability criterion and fronts are usually referred to as pulled \cite{VanSaarloos03}. In the opposite case, the asymptotic speed can only be determined by nonlinear methods and fronts are referred to as pushed \cite{VanSaarloos03}. Variational methods are particularly efficient to determine appropriate approaches for the minimum speed for pushed fronts \cite{Benguria}. FKPP fronts have been observed in Taylor-Couette \cite{Ahlers}, Rayleigh-Benard experiments \cite{Fineberg}, pearling and pinching on the propagating Rayleigh instability \cite{Powers}, spinodal decomposition in polymer mixtures \cite{Langer92} and liquid crystal light valve \cite{Residori04}, to mention a few.

The above scenario changes completely for a front connecting two stable states. In the case of two uniform states, a gradient system tends to develop the most stable state, in order to minimize its energy, so that the front always propagates toward the most energetically favorable state. It exists only as one point in parameter space for which the front is motionless, which is usually called the Maxwell point, and is the point for which the two states have exactly the same energy \cite{Goldstein}. The evolution of front solutions can be regarded as a particle-type one, i.e., they can be characterized by a set of continuous parameters such as position, width of core and so forth. Close to the Maxwell point, based on the method of variation of parameters, one can analytically determine the speed of the front \cite{Pomeau}. For variational systems away from the Maxwell point one can have implicit expressions for the front speed \cite{Pomeau}. In non-variational case, the analytical expression of the front speed is a problem still unresolved.

System with discrete reflection symmetry possess two equivalent states with interfaces, or domain walls which are generically at rest. Indeed, the two connected states are "energetically" equivalent. These front solutions are termed kinks. However, under spontaneous breaking of the parity symmetry, these fronts can acquire a nonzero asymptotic speed. This phenomenon is the denominated Ising-Bloch transition \cite{Coullet90} and has been observed in systems as different as  ferromagnetic systems \cite{Bulaevsky,Clerc09}, liquid crystals \cite{Gilli,Kawagishi}, chemical reactions \cite{Haim}, and nonlinear optical cavities \cite{Esteban-Martin}. Gradient or variational systems do not exhibit this phenomenon, because the front speed is proportional to the energy difference between the two equivalent states. 
A different situation is that of a front connecting a homogeneous and a periodic state. Indeed, the existence of a pinning range for a front between these states was predicted by Pomeau \cite{Pomeau}. In this case, a pinning-depinning phenomenon is expected to occur as a result of the competition between a symmetry breaking of the global energy that tends to favor the front propagation in one direction and spatial modulations that tend to block the front by introducing local potential barriers at the front core dynamics \cite{Bensimon,Clerc-Falcon}. Depending on the dominating effect, the front can either stay motionless- blocked over a large range, therefore called the pinning range-or propagate with periodic leaps apart from it. Starting from a critical value of the control parameter, the pinning-depinning transition occurs by a loss of stability of the pinned front.

The existence of a pinning range, and associated pinning-depinning phenomenon, has a fundamental relevance in numerous domains where front propagation is involved. As examples, we can cite vibrated fluids \cite{Epstein}, chemical reactions \cite{Petrov97,Schwartz}, microfluidic chips \cite{Thiam}, wetting of micro-structured surfaces \cite{Sbragaglia}, control of the motility of bacteria \cite{Douarche}, and growth of self-assembly monolayers \cite{Douglas}. Indeed, spatial discreteness and spatial modulations can be seen as the bases for a wealth of life behaviors, where the emergence of complexity results from the microscopic granularity of the system \cite{Shnerb}. As for the experimental studies of the pinning-depinning phenomenon, only a few approaches have been proposed up to date. In a two-dimensional spatially forced system, an experimental characterization of front propagation has shown the anisotropy of the front velocity \cite{Armero}; however, the issue of a pinning range was not addressed. Only recently, by employing a 1D spatially forced liquid-crystal system, the experimental demonstration of the existence of a pinning range has been achieved \cite{Haudin}.

\section{Temporal localized structures}
In the temporal domain, localized structures have also been observed in passive fiber resonators. These devices constitute a basic configuration in nonlinear fiber optics.  More specifically, experimental studies have demonstrated that when these resonators are pumped by a continuous wave, they exhibit spontaneously self-organized temporal structures in the form of trains of short pulses with well defined repetition rate \cite{Mitschke,COEN-Hal,Stratmann}. The theoretical description of all fiber resonators is provided by the Lugiato and Lefever model (LL model, \cite{LuLe}). The breakup of continuous wave into trains of pulses is attributed to the competition between the  nonlinear mechanism which is originated from the intensity-dependent refractive index  (Kerr effect) that tends to amplify locally the field intensity, the chromatic dispersion which on the contrary tends to restore uniformity, and dissipation.

Besides a  pulse train distribution, temporal localized structures (TLS) are found in a well-defined region of parameters called a pinning zone. In this regime, the system exhibits a coexistence between two states: the uniform background and the train of pulses of light that emerges from subcritical modulational instability \cite{Scroggie}. They have been experimentally observed in fiber ring resonator \cite{Leo}. Theoretical predictions and experimental observations of temporal localized structures using standard optical fibers allow for the conception of all-optic systems for generation of signals with high repetition rate or to operate as all-optical memories with capacity that can reach  25 Gbits/s \cite{Leo}.  In addition to temporal localised structures, front dynamics have been analytically predicted and experimentally observed in a simple passive fiber cavity synchronously pumped by a pulsed laser \cite{convection-fiber}. Synchronous pumping means that the time-of-flight of the light pulses in the cavity is adjusted to the laser repetition time. Recently, an asymptotic behavior of the front velocity has been established near the up-switching point of the bistable response curve \cite{Coulibaly-TT}.  The front dynamics obeys a generic power law when the front velocity approaches asymptotically its linear growing value as predicted by van Saarloos \cite{VanSaarloos03}.

When all fiber cavity is operating close to the zero dispersion wavelength, high-order chromatic dispersion effects may play an important role in the dynamics of a photonic crystal fiber (PCF) \cite{Cavalcanti91,Pitois03-2,joly05}. PCFs allow for a high control of the dispersion curve  and permit exploring previously inaccessible parameter regimes \cite{Russel,Schmidberger12}. The inclusion of the fourth order dispersion allows the modulational instability to have a finite domain of existence delimited by two pump power values \cite{Tlidi1} for the stabilization of dark temporal cavity solitons in PCF resonators \cite{Tlidi-Gelens}. Together with this effect,  the third order dispersion causes a spontaneous broken reflection symmetry and allows  the motion of both  periodic and localized structures that are generated in the resonator \cite{Tlidi-Lyes}. 

\section{Vegetation pattern in semi(arid) ecosystems}
Desertification is one of the grand environmental challenges in EU and in the world. In last decade, the concepts of symmetry breaking and self-organization have been applied to plant ecology systems where dissipative structures are generically called vegetation patterns. A well-documented example is "tiger bush". The observation of this phenomenon was possible thanks to the development of aerial photography in the early forties. Vegetation patterns were described for the first time 64 years ago by Macfadyen in British Somaliland \cite{Macfadyen}. In this paper, Macfadyen states that:

{\it{The study of the first two patterns, of which I have been able to find no previous recognition, is not conveniently pigeon-holed under any one accepted branch of knowledge, and the phenomena are thus awkward to classify. They are manifestly within the province of botany and ecology; the essential background concerns geomorphology and meteorology; the causes as I believe, must be investigated by physics and mathematics. \cite{Macfadyen}}}

Since then, vegetation patterns  have been observed and studied in many regions throughout the world \cite{White}. Vegetation patterns are sparsely populated or bare areas alternating with dense vegetation patches. They are made of either  grass or trees and shrubs. By increasing the aridity, the spatial distribution  (periodic or aperiodic) of patches first displays bare spots, then stripes, and finally spots. The spatial distribution occurs on a decametric to hectometric scale. This large scale phytosociological self-organization leading to nonuniform and non random distribution of the phytomass is the rule rather than the exception. They were observed on vast territories of many arid and semiarid regions of Africa, Middle East, Australia and Noth America. The terms arid and semiarid refer to climatic conditions where the water and nutrient resources are scarce; typically, in regions where the annual rainfall ($50-750$ mm) is low compared to the potential of evapo-transpiration (PET>$1.5 \times 10^3$ mm).  

Mathematical modeling of the vegetation growth constitutes an important tool towards the understanding of the mechanism leading to  the desertification phenomenon in semiarid and arid landscapes. To explain the origin of the vegetation patterns three main mechanisms have been proposed.  The following classification of ecological models was suggested by Lefever and Turner in \cite{Lefever-Turner}:\\
(i) The non-local FKPP (Fisher, Kolmogorov, Petrovsky, Piscounov) approach: it focuses on the relationship between the structure of individual plants and the facilitation-competition interactions existing within plant communities. The biomass density is defined at the individual plant level; the modeling calls for no other state variable \cite{LL,LT,Lefever_2009,Tli-Lef-Vla}.
\\
(ii) The reaction-diffusion approach: it emphasizes the influence on vegetation patches of water transport by below ground
diffusion and/or above ground run-off. The biomass density is defined at the patch level. Together with the water
concentration below ground and/or in some surface ground layer, it constitutes the set of state variables \cite{Klausmeir,Sherartt,Rietkerk_2001,Okayasu,Meron2004,Gowda}.
\\
(iii) The stochastic approach: it focuses on the "constructive" role of environmental randomness as a source of noise induced
symmetry breaking transitions triggering pattern formation. These transitions occur when the variance of environmental
noise increases while its average value remains constant \cite{Odorico1,Rev17}.

Besides the behavior of periodic vegetation patterns, the same self-organization mechanism predicts the existence of aperiodic, localized structures. They  consist of localized patches of vegetation, randomly distributed on bare soil~
\cite{Lejeune-Tlidi-Couteron,RIETKERK4,Meron2001} or, on the contrary, consist of localized spots of bare soil, randomly distributed in an otherwise uniform vegetation cover \cite{Tli-Lef-Vla}. Well documented examples are the fairy circles existing in the Namib desert.  These circular areas devoid of any vegetation can reach diameters of up to 14 m, which exceeds by more than one order of magnitude the size of the tall grasses surrounding them. In a recent study, van Rooyen {\it et al.}~\cite{ROOYEN} have investigated in depth the strengths and shortcomings of several hypotheses concerning their origin. In this respect, the authors have been able to rule out external causes such as a possible existence of localized radioactive areas unsuited for the development of plants or a link between fairy circles and the activity of termites or the proximity of their nests; among the other hypotheses considered by the authors (notably, the release of allelopathic compounds and the possible existence of stimulatory or inhibitory influences due to interactions between different plant species), none was found which fitted satisfactorily their experimental observations. Recent progress has been made in the modeling of fairy circles \cite{Tli-Lef-Vla,Jankowitz2008,Tschinkel2012,Cramer2013,Juergens2013,Getzin2014}. The origin of this phenomenom is still matter of debate.

\section{Contributions to the theme issue}
The contributions to this theme issue are devoted to the investigation of the formation, stability and control of localized structures in various research areas: nonlinear optics (including laser dynamics and fiber optics), fluid mechanics, chemistry and plant ecology. The results presented in the contributions indicate that considerable progress has been made in understanding of the properties of non-linear localised structures in dissipative systems. Most of contributions deal with  non-linear photonics devices having potential applications for all-optical control of light, optical storage, and information
 processing.  The contributions are grouped by research area. The paper by Vladimirov et al. \cite{VladimirovPTRS} presents the first experimental evidence of localized structures of light in a medium size bottom-emitting InGaAs multiple quantum well VCSEL operated in injection-locked regime. A scheme based on delay feedback to control the self-mobility of such localized structures  was proposed and studied theoretically. In the case of two coupled VCSELs, Turconi et al. \cite{BarlandPTRS} perform a statistical analysis of time-resolved commutation experiments  and investigate the role of wavelength, polarization and pulse energy in the switching process. Vahed et al. \cite{PratiPTRS} investigate periodic and chaotic localized structures in VCSEL with a saturable absorber.    Using a nematic liquid crystal cells with a photosensitive wall and negative dielectric anisotropy,  Barboza et al \cite{ResidoriPTRS} demonstrate experimentally and numerically the possibility of generation of stable vortices with swirling arms. The demonstration and the characterization  of families of moving solitons in   lattices composed of metallic nanoparticles with nonlinear Kerr-like response and an external driving field is provided by  Noskov et al. \cite{KivsharPTRS}.  A system formed by  an array of  plasmonic waveguides, with gain applied to selected hot spot cores  is discussed by Ding et al. \cite{MalomedPTRS}. A review on localized modes in dissipative lattice media is presented by  Malomed et al. \cite{MihalachePTRS}.  Another review 
on the dynamics of particles, waves as well as spatial and temporal localized structures in cavities formed by oscillating mirrors is contributed by  Rosanov \cite{RosanovPTRS}.  Various collisions of 2D-localized structures are investigated and a possibility of breather formation as a result of collision is presented. The contribution by de Varcarcel  et al.  \cite{DevarcarcelPTRS} presents a review on rocking phenomenon in various spatially extended systems and discuss their applications in optics. A Nagumo-Kuramoto model exhibiting a  localised spatiotemporal chaotic type of behavior often called chaoticon is discussed by Verschueren et al. \cite{ClercPTRS}. The mechanism responsible for the appearance of such chaotic regime is attributed to front interactions.  Tesio et al. \cite{FirthPTRS} discuss the self-organisation phenomenon in cold atomic gases that manifests itself as a spontaneous emergence of long-range spatial order from fluctuations in the transverse plane. Using the Kuramoto model, this process is interpreted as a synchronisation transition in a fully connected network of fictitious oscillators. In nonlinear fiber optics,  a  vector model describing an erbium doped mode-locked fibre laser with carbon nanotube saturable absorber is investigated by Sergeyev \cite{SergeyevPTRS}. In particular, the author discusses the formation of a new family of vector dissipative solitons with fast and slowly evolving polarization states in this system. Bahloul et al. \cite{BahloulPTRS} perform analytical and numerical studies of  the combined influence of third and fourth order dispersion on the dynamics of temporal localized structures in photonic crystal fiber cavity. This analysis  reveals the existence of  moving temporal localized structures of light.   

Chabchoub et al.  \cite{AkhmedievPTRS} provide for the first time evidence of  localised structures in the form of  Kuznetsov-Ma soliton and  Akhmediev breather  in a water wave flume. Their analytical study based on  nonlinear Schr\"odinger equation agrees with the experimental observations. The control of  breathing localized structures by time-delayed optical feedback in a three-component reaction-diffusion system is investigated theoretically by Gurevich \cite{GurevichPTRS}. It is shown that varying the delay parameters can lead either to stabilization of the breathing or to delay-induced periodic and quasiperiodic oscillations of localized structures.  Finally,  three contributions on localized patches, fairy circles, and dissipative structures in plant ecology are presented in the end of the issue. They can be summarized as follows. The first contribution by Couteron et al. \cite{CouteronPTRS} presents plant morphologies observed in the Andean altiplano and interpret them in terms of localized structures  leading to a patchy, aperiodic distribution of the vegetation cover.  The second paper by Fernandez-Oto et al. \cite{OTOPTRS} demonstrates that the enigmatic fairy circles of the Namibia desert can be stabilized  by applying a strong nonlocal coupling which models the competition process between the plants. Numerical and analytical results obtained by the authors are in close agreement with the field observations. The last paper by Martinez-Garcia et al. \cite{LopezPTRS} discusses the formation of vegetation patterns in arid and semi arid landscapes. They attribute this formation to the competition between plants and not to the facilitation phenomenon.

\section*{Acknowledgment}

M.G.C. acknowledge the financial support of FONDECYT project 1120320. C.F.O.  the financial support of   Becas Chile. M .T received support from the Fonds National de la Recherche Scientifique (Belgium). This research was supported by the Interuniversity Attraction Poles program of the Belgian Science Policy Office, under grant IAP P7-35.

\end{document}